\pdfoutput=1
\documentclass[11pt]{article}

\usepackage{EMNLP2022}

\usepackage{times}
\usepackage{latexsym}
\usepackage{multirow}
\usepackage{graphicx}
\usepackage{booktabs}
\usepackage{ulem}
\usepackage[T1]{fontenc}
\usepackage[utf8]{inputenc}

\usepackage{microtype}
\usepackage{amsmath}
\usepackage{amssymb}
\usepackage{ulem}
\usepackage{soul}
\soulregister\cite7
\newtheorem{theorem}{Theorem}
\usepackage{xspace}
\usepackage{graphics}
\usepackage{arydshln}
\usepackage{inconsolata}

\newcommand{\Sec}{Section\xspace}

\title{Exploring Representation-Level Augmentation for Code Search}

\author{{\bf Haochen Li}$^{1}$\ \ \ \ 
    {\bf Chunyan Miao}$^{1,2}$\thanks{\ \ Corresponding author.}\ \ \ \ 
    {\bf Cyril Leung}$^{1,2}$ \ \ \ \ 
  {\bf Yanxian Huang}$^3$\ \ \ \ \\
  {\bf Yuan Huang}$^3$\ \ \ \
  {\bf Hongyu Zhang}$^4$\ \ \ \
  {\bf Yanlin Wang}$^3$\\
    $^1$School of Computer Science and Engineering, Nanyang Technological University, Singapore \\
    $^2$China-Singapore International Joint Research Institute (CSIJRI), China \\
    $^3$School of Software Engineering, Sun Yat-sen University, China \\
    $^4$The University of Newcastle, Australia \\
     \texttt{\{haochen003,ascymiao,cleung\}@ntu.edu.sg, huangyx353@mail2.sysu.edu.cn} \ \\\texttt{\{huangyuan5,wangylin36\}@mail.sysu.edu.cn, hongyu.zhang@newcastle.edu.au}
     }

\begin{document}
\maketitle

\begin{abstract}
Code search, which aims at retrieving the most relevant code fragment for a given natural language query, is a common activity in software development practice. Recently, contrastive learning is widely used in code search research, where many data augmentation approaches for source code (e.g., semantic-preserving program transformation) are proposed to learn better representations.  However, these augmentations are at the raw-data level, which requires additional code analysis in the preprocessing stage and additional training costs in the training stage.    
In this paper, we explore augmentation methods that augment data (both code and query) at representation level which does not require additional data processing and training, and based on this we propose a general format of representation-level augmentation that unifies existing methods. Then, we propose three new augmentation methods (linear extrapolation, binary interpolation, and Gaussian scaling) based on the general format. Furthermore, we theoretically analyze the advantages of the proposed augmentation methods over traditional contrastive learning methods on code search. 
We experimentally evaluate the proposed representation-level augmentation methods with state-of-the-art code search models on a large-scale public dataset consisting of six programming languages. The experimental results show that our approach can consistently boost the performance of the studied code search models. Our source code is available at \url{https://github.com/Alex-HaochenLi/RACS}.
\end{abstract}

\section{Introduction}

In software development, developers often search and reuse commonly used functionalities to improve their productivity \cite{nie2016query,shuai2020improving}. With the growing size of large-scale codebases such as GitHub, retrieving semantically relevant code fragments accurately becomes increasingly important in this field~\cite{allamanis2018survey,liu2021opportunities}.

Traditional approaches~\cite{nie2016query, yang2017iecs, rosario2000latent, hill2011improving, satter2016search, lv2015codehow,van2017combining} leverage information retrieval techniques to treat code snippets as natural language text and match certain terms in code with queries, hence suffering from the vocabulary mismatch problem \cite{mcmillan2011portfolio, robertson1995okapi}. Deep siamese neural networks first embed queries and code fragments into a joint embedding space, then measure similarity by calculating dot product or cosine distance \cite{lv2015codehow,cambronero2019deep,gu2021multimodal}. Recently, with the popularity of large scale pre-training techniques, some big models for source code  \cite{guo2020graphcodebert,feng2020codebert,guo2022unixcoder,wang2021syncobert, jain2020contrastive, li2022coderetriever} with various pre-training tasks are proposed and significantly outperform previous models.

Contrastive learning is widely adopted by the above-mentioned models. It is suitable for code search because the learning objective aims to push apart negative query-code pairs and pull together positive pairs at the same time.  
In contrastive learning, negative pairs are usually generated by In-Batch Augmentation \cite{huang2021cosqa}. 
For positive pairs, besides labeled ones, some researchers proposed augmentation approaches to generate more positive pairs
\cite{bui2021self, jain2020contrastive, fang2020cert,simcse, moco}. 
The main hypothesis behind these approaches is that augmentations do not change the original semantics. However, these approaches are resource-consuming \cite{batchmixup,jeong2022augmenting}. Models have to embed the data again for the augmented data.

To solve this problem, some researchers proposed representation-level augmentation, which augments the representations of the original data. For example, linear interpolation, a representation-level augmentation method, is adopted by many classification tasks in NLP~\cite{guo2019augmenting, DBLP:conf/coling/SunXYLYH20, DBLP:conf/emnlp/Du0WX0LJ21}. 
The augmented representation  
captures the structure of the data manifold and hence could force model to learn better features, as argued by \citet{DBLP:conf/icml/VermaLKPL21}. These augmentation approaches are also considered to be semantic-preserving.

The representation-level augmentation methods are not investigated on the code search task before. 
To the best of our knowledge, \citet{jeong2022augmenting} is the only work to bring representation-level augmentation approaches to a retrieval task. Besides linear interpolation, it also proposes another approach called stochastic perturbation for document retrieval. Although these augmentation methods bring improvements to model performance, they are not yet fully investigated. The relationships between the existing methods and how they affect model performance remain to be explored.

In this work, we first unify linear interpolation and stochastic perturbation into a general format of representation-level augmentation. We further propose three  augmentation methods (linear extrapolation, binary interpolation, and Gaussian scaling) based on the general format. Then we theoretically analyze the advantages of the proposed augmentation methods based on the most commonly used InfoNCE loss \cite{van2018representation}. As optimizing InfoNCE loss equals to maximizing the mutual information between positive pairs, applying representation-level augmentation leads to tighter lower bounds of mutual information. We evaluate representation-level augmentation on several Siamese networks across several large-scale datasets. Experimental results show the effectiveness of the representation-level augmentation methods in boosting the performance of these code search models.  
To verify the generalization ability of our method to other tasks, we also conduct experiments on the paragraph retrieval task, and the results show that our method can also improve the performance of several paragraph retrieval models.

In summary, our contributions of this work are as follows:
\begin{itemize}
    \item We unify previous representation-level augmentation methods to propose a general format. 
    Based on this general format, we propose three novel augmentation methods.
    \item We conduct theoretical analysis to show that representation-level augmentation has tighter lower bounds of mutual information between positive pairs.
    \item We apply representation-level augmentation on several code search models and evaluate them on the public CodeSearchNet dataset with six programming languages. Improvement of MRR (Mean Reciprocal Rank) demonstrates the effectiveness of the representation-level augmentation methods.
\end{itemize}

The rest of the paper is organized as follows. We introduce related work of code search and data augmentation in \Sec~\ref{sec:rw}. \Sec~\ref{sec:approach} introduces the main part, including the general format of representation-level augmentation, new augmentation methods, and their application on code search. In \Sec~\ref{sec:theory}, we analyze the theoretical lower bounds of mutual information and study why our approach works. In \Sec~\ref{sec:expsetup} and \Sec~\ref{sec:results}, we conduct extensive experiments to show the effectiveness of our approach. Then we discuss the generality of our approach in \Sec~\ref{sec:discussion} and \Sec~\ref{sec:conclusion} concludes this paper.

\section{Related Work}\label{sec:rw}

\subsection{Code search} 
As code search can significantly improve the productivity of software developers by reusing functionalities in large codebases, finding the semantic-relevant code fragments precisely is one of the key challenges in code search. 

Traditional approaches leverage information retrieval techniques that try to match some keywords between queries and codes \cite{mcmillan2011portfolio, robertson1995okapi}. These approaches suffer from vocabulary mismatch problem where models fail to retrieve the relevant codes due to the difference in semantics. 

Later, deep neural models for code search are proposed. They could be divided into two categories, early fusion and late fusion. Late fusion approaches \cite{gu2018deep,husain2019codesearchnet} use a siamese network to embed queries and codes into a shared vector space separately, then calculate dot product or cosine distance to measure the semantic similarity. Recently, following the idea of late fusion, some transformer-based models with specifically designed pre-training tasks are proposed \cite{feng2020codebert,guo2020graphcodebert,guo2022unixcoder}. They significantly outperform previous models by improving the understanding of code semantics. Instead of calculating representations of queries and codes independently, early fusion approaches model the correlations between queries and codes during the embedding process \cite{li2020learning}. \citet{li2020learning} argues that early fusion makes it easier to capture implicit similarities. For applications of an online code search system, late fusion approach facilitates the use of neural models because the code representations can be calculated and stored in advance. During run time, only query representations need to be computed. Thus, in this work, we focus on late fusion approaches.

\subsection{Data augmentation} 
Data augmentation has long been considered crucial for learning better representations in contrastive learning. The augmented data are considered to have the same semantics with the original data. For the augmentation of queries, synonym replacement, random insertion, random swap, random deletion, back-translation, spans technique and word perturbation can be potentially used to generate individual augmentations \cite{wei2019eda,giorgi2020declutr,fang2020cert}. For the augmentation of code fragments, \citet{bui2021self} proposed six semantic-preserving transformations: {\it Variable Renaming}, {\it Permute Statement}, {\it Unused Statement}, {\it Loop Exchange}, {\it Switch to If} and {\it Boolean Exchange}. These query and code augmentation approaches have one thing in common, that is, the transformation is applied to the original input data. 
Another category is augmenting during the embedding process. Models can generate different representations of the same data by leveraging time-varying mechanisms. MoCo \cite{moco} encodes data twice by the same model with different parameters. SimCSE \cite{simcse} leverages the property of dropout layers by randomly deactivating different neurons for the same input. Methods described in this paragraph are resource-consuming because models embed twice to get representations of original data and augmented one.

For representation-level augmentation on NLP tasks, linear interpolation is widely used on classification tasks in previous work \cite{guo2019augmenting, DBLP:conf/coling/SunXYLYH20, DBLP:conf/emnlp/Du0WX0LJ21}. They take the interpolation result as noised data and want models to classify the noised one into the original class. \citet{DBLP:conf/icml/VermaLKPL21} theoretically analyzed how interpolation noise benefits classification tasks and why it is better than Gaussian noise.  \citet{jeong2022augmenting} is the first to introduce linear interpolation and perturbation to the  document retrieval task. However, the effect and intrinsic relationship of these two methods are not fully investigated.

\section{Approach}\label{sec:approach}
In this section, we unify the linear interpolation and stochastic perturbation into a general format. Based on it, we propose three other augmentation methods for the code retrieval task, linear extrapolation, binary interpolation and Gaussian scaling. Then, we explain how to apply representation-level augmentation with InfoNCE loss in code retrieval.

\subsection{General format of representation-level augmentation}

For simplicity, we take code augmentation as an example to elaborate the details. The calculation process is similar when applying to query augmentations. Given a data distribution $\mathbf{D}=\{x_i\}_{i=1}^K$ where $x_i$ is a code snippet, $K$ is the size of the dataset. We use an encoder function $h: \mathbf{D} \rightarrow \mathbf{H}$ to map codes to representations $\mathbf{H}$.

\paragraph{Linear interpolation} 
Linear interpolation randomly interpolate $h_i$ with another chosen sample $h_j$ from $\mathbf{H}$:
\begin{equation}
\label{linearinterpolate}
    h_i^+ = \lambda h_i + (1-\lambda)h_j
\end{equation}

where $\lambda$ is a coefficient sampled from a random distribution. For example, $\lambda$ can be sampled from a uniform distribution $\lambda \sim U(\alpha , 1.0)$ with high values of $\alpha$ to make sure that the augmented data has similar semantics with the original code $x_i$.

\paragraph{Stochastic perturbation} 
Stochastic perturbation aims at randomly deactivating some features of representation vectors. In order to do so, masks are sampled from a Bernoulli distribution $B(e, p)$, where $e$ is the embedding dimension. $p$ is a low probability value since we only deactivate a small proportion of features. For implementation, we could use Dropout layers.

\paragraph{General format of representation-level augmentation} 
We revisit the above two augmentation approaches and unify them into a general format, which could be described as:

\begin{equation}
\label{generalformat}
    h^+ = \alpha \odot h + \beta \odot h^\prime
\end{equation}

where $h,h^\prime \in \mathbf{H}$, $\alpha$ and $\beta$ are coefficient vectors. For linear interpolation, $\alpha = \lambda$,  
$\beta = 1-\lambda$, $h, h^\prime \in \mathbf{H}$, and $ h \neq h^\prime$. For stochastic perturbation, $\alpha \in \{0, \frac{1}{1-p} \}^e$ where elements of $\alpha$ are sampled from a Bernoulli distribution $B(p)$, $\beta = \frac{1}{1-p}-\alpha$, $h \in \mathbf{H}$, and $h^\prime = 0$. 

\subsection{New augmentation methods}
Based on the general format, we provide three new
augmentation methods for the code retrieval task.

\paragraph{Binary interpolation}
Binary interpolation randomly swaps some features with another chosen sample. The difference between binary interpolation and stochastic perturbation is that the former swaps with other samples while the latter swaps with zero vector. Specifically, for binary interpolation, $\alpha \in \{0,1\}^e$ where elements of $\alpha$ are sampled from a Bernoulli distribution $B(p)$, $\beta = 1-\alpha$, $h, h^\prime \in \mathbf{H}$, and $ h \neq h^\prime$.

\paragraph{Linear extrapolation}
\citet{wang2020understanding} concludes that optimizing contrastive loss makes feature vectors roughly uniformly distributed on the hypersphere. As linear interpolation generates augmented data inside the hypersphere, linear extrapolation oppositely generates outside ones. $\lambda$ is sampled from a uniform distribution $\lambda \sim U(1.0, \alpha)$ with small values of $\alpha$. Other settings are the same as linear interpolation.

\paragraph{Gaussian scaling}
Gaussian scaling generates scaling coefficients for each feature in the representation, which can be considered as a type of perturbation noise.  Compared with directly adding Gaussian noise, the proposed scaling noise captures the structure of the data manifold, which may force networks to learn better representations. If we describe Gaussian scaling in general format, then $\alpha = 1$, $\beta \sim N(0, \sigma)$ with small values of $\sigma$, $h=h^\prime \in \mathbf{H}$.

\subsection{Contrastive learning with representation-level augmentation for code search} 

Contrastive learning seeks to satisfy that similarities between positive pairs are greater than that between negative pairs, which is suitable for code search. In previous works, in order to optimize the objective, several loss functions are proposed, including triplet loss \cite{triplet}, max-margin loss \cite{hadsell2006dimensionality}, and logistic loss \cite{logistic}. In this work, we consider InfoNCE loss \cite{van2018representation} because of its better performance hence dominant adaption in current contrastive learning models. For the effect of representation-level augmentation on other loss functions, we empirically analyze it in Appendix \ref{otherloss}.

We start from the vanilla InfoNCE loss. Suppose we have a batch of $B$ samples consisting of queries and codes. We encode queries and codes to get query representations $Q=\{q_i \}_{i=1}^B$ and code representations $C=\{c_j \}_{j=1}^B$ using the encoder function $h$. $(q_i,c_{j})$ are positive pairs when $i=j$ and negative pairs otherwise. Therefore, for each query, we could generate 1 positive pair and $B-1$ negative pairs. The InfoNCE loss tries to maximize the similarity between one positive pair and minimize the similarity between other negative pairs. Here we use dot product as the measurement of similarity and in \Sec~\ref{cos} we will discuss the effect of using dot product compared to cosine distance. The loss can be described as:
\begin{equation}
\label{infonce}
    L = -\mathbb{E} \left [\log \frac{\exp (q_i\cdot c_i)}{\exp (q_i\cdot c_i)+\sum_{j\neq i}^B\exp (q_i \cdot c_j)} \right]
\end{equation}

Then, we conduct representation-level augmentation. We randomly choose one out of the five augmentation methods and augment the original queries $Q$ and original codes $C$ for $N$ times. In each augmentation, the augmentation approach is fixed, but the coefficients $\alpha$ and $\beta$ are randomized hence different. After augmentation, we get augmented query and code sets, $Q^+ = \{ q_{ni}\}_{n=1,i=1}^{n=N,i=B}$ and $C^+ = \{ c_{nj}\}_{n=1,j=1}^{n=N,j=B}$. We follow the hypothesis of other representation-level augmentation methods that the augmented representation still preserves or is similar to 
% \yl{revised here, please check} 
the original semantics. Therefore, for a certain query $q_i$, we consider $q_i$ and $q_{ni}$,$\forall n \in [1,N]$ share the same semantic meaning. And this similarly applies to $c_j$ and $c_{nj}$, $\forall n \in [1,N]$. Since $(q_i,c_j)$ is labeled as a positive pair when $i=j$, $\forall n \in [1,N]$ $(q_i, c_{nj})$ and $(q_{ni}, c_j)$ are also naturally labeled as positive pairs. Similarly, we get $(q_{ni}, c_{nj,j \neq i})$ as negative pairs.
Thus, compared with vanilla InfoNCE loss, we now have $(N+1)^2B$ positive pairs in total and for each query $q \in Q\cup Q^+$ we can generate $(B-1)(N+1)$ negative pairs. 

\section{Theoretical Analysis}\label{sec:theory}
In this section, we mathematically analyze the effect of InfoNCE loss with or without representation-level augmentation and prove that optimizing InfoNCE loss with representation-level augmentation leads to mutual information with tighter lower bounds between positive pairs. Here we take linear interpolation as an example to demonstrate the benefits. Other forms of representation-level augmentation are left to future work. The mutual information between a query $q$ and a code fragment $c$ is:

\begin{equation}
\label{mi}
    I(q,c)=\mathbb{E}_{q,c}\left [ \log \frac{p(c|q)}{p(c)} \right ]
\end{equation}

\begin{theorem}
\label{theorem1}
 Optimizing InfoNCE loss $\mathcal{L_N}$ improves lower bounds of mutual information $I(q, c)$ for a positive pair:
 
 \begin{equation}
 \label{theory1}
    I(q,c) \ge \log (B)-\mathcal{L_N}
 \end{equation}
 
 where $q \in Q$, $c \in C$, and $B$ is the size of sets.
\end{theorem}

Proof of Theorem \ref{theorem1} is presented in Appendix \ref{appendix:theory1}. This is proved in the original paper of InfoNCE loss \cite{van2018representation}. Here, to better extend the proof of Theorem \ref{theorem2}, we prove it in another way.

\begin{theorem}
\label{theorem2}
Optimizing InfoNCE loss $\mathcal{L_N}$ with representation-level augmentation improves a tighter lower bounds of mutual information $I(q, c)$ for a positive pair:

\begin{equation}
\label{theory2}
\begin{split}
    I(q,c) \ge &  \frac{1}{\alpha^2} ( \log (NB)-\mathcal{L_N} \\
    & -\alpha\beta\cdot I(q,c^-)-\alpha\beta\cdot I(q^-,c) \\
    & -\beta^2\cdot I(q^-, c^-) )
\end{split}
\end{equation}

where $q,q^-\in Q$,$c,c^-\in C$, $(q,c^-)$, $(q^-,c)$ and $(q^-, c^-)$ are all negative pairs, $\alpha$ and $\beta$ are coefficients in Equation \ref{generalformat}, $B$ is the size of sets, and $N$ is the augmentation time.
\end{theorem}

Proof of Theorem \ref{theorem2} is presented in Appendix \ref{appendix:theory2}. Since we interpolate $q$ with other samples $q^-$ in the batch ($c$ with $c^-$), the mutual information between $(q,c^-)$, $(q^-,c)$ and $(q^-, c^-)$ are also incorporated into the optimizing process. As defined in Eq.\ref{linearinterpolate}, $\beta$ is a small value that close to 0. According to Eq.\ref{mi}, for negative pairs $p(c|q)$ can be expressed as $\frac{p(c)p(q)}{p(q)}$ due to the independence of sampling $q$ and $c$. Considering this, we can see that the optimal mutual information between negative pairs is also 0. Note that we interpolate $q$ and $c$ with different samples to make sure that $(q^-,c^-)$ is a negative pair. Thus, the last three terms in Eq.\ref{theory2} can be ignored. Comparing $\frac{1}{\alpha^2}(\log (NB)-\mathcal{L_N})$ with $\log (B)-\mathcal{L_N}$, we can see that representation-level augmentation improves the lower bounds of mutual information.

\section{Experimental Setup}\label{sec:expsetup}
In this section, we describe datasets, baselines, and implementation details.

\begin{table}[t]
\centering 
% \small
\setlength{\tabcolsep}{2.5pt}
% \resizebox{\columnwidth}{!}{%
\begin{tabular}{lcccc}
\toprule
Language & Training & Validation & Test & Codebase \\ \midrule
Ruby & 24,927 & 1,400 & 1,261 & 4,360 \\
JavaScript & 58,025 & 3,885 & 3,291 & 13,981 \\
Go & 167,288 & 7,325 & 8,122 & 28,120 \\
Python & 251,820 & 13,914 & 14,918 & 43,827 \\
Java & 164,923 & 5,183 & 10,955 & 40,347 \\
PHP & 241,241 & 12,982 & 14,014 & 52,660 \\ \bottomrule
\end{tabular}%
% }
\caption{CodeSearchNet dataset statistics.}
\label{datastat}
\end{table}

\subsection{Datasets}
To evaluate the effectiveness of representation-level augmentation, we use a large-scale benchmark dataset \textbf{CodeSearchNet} (CSN) \cite{husain2019codesearchnet} that is widely used in previous studies \cite{guo2020graphcodebert,feng2020codebert,guo2022unixcoder}. It contains six programming languages including Ruby, JavaScript, Go, Python, Java, and PHP. The statistics of the dataset are shown in Table \ref{datastat}. For the training set, it contains positive-only query-code pairs. For validation and test sets, they only have queries and the model retrieves the correct code fragments from a fixed codebase. Here we follow \cite{guo2020graphcodebert} to filter out low-quality examples (such as code that cannot be successfully parsed into Abstract Syntax Trees).

We measure the performance using Mean Reciprocal Rank (MRR) which is widely adopted in previous studies. MRR is the average of reciprocal ranks of a true code fragment for a given query $Q$. It is calculated as:
\begin{equation}
    MRR=\frac{1}{|Q|}\sum_{i=1}^{|Q|}\frac{1}{Rank_i}
\end{equation}
where $Rank_i$ is the rank of the correct code fragment that is related to the i-th query.

\subsection{Baselines}
Since representation-level augmentation is orthogonal to siamese networks, we apply it to several models:
\begin{itemize}
    \item \textbf{ RoBERTa (code)} is pre-trained with mask language modeling (MLM) task on code corpus \cite{husain2019codesearchnet}.
    \item \textbf{ CodeBERT} is a bi-modal pre-trained model pre-trained on two tasks: MLM and replaced token detection \cite{feng2020codebert}. Note that in this work we refer CodeBERT to the siamese network architecture described in the appendix of the original paper. 
    \item \textbf{ GraphCodeBERT} takes the structure information of codes into account. \cite{guo2020graphcodebert} develops two structure-based pre-training tasks: data flow edge prediction and node alignment. 
    \item \textbf{ UniXCoder} leverages cross-model contents like AST and comments to enhance code representation \cite{guo2022unixcoder}.
\end{itemize}

\subsection{Implementation details} 
For all the settings of these models except the training epoch, we follow the original paper. For representation-level augmentation, since linear extrapolation and linear interpolation is similar, we implement it as one approach. During training, we randomly choose one augmentation approach out of four with equal probability for a batch and augment data 5 times. We re-sample data and coefficients to augment the original data in each augmentation. Specifically, for linear interpolation and extrapolation, we sample $\alpha$ from a uniform distribution $U \sim (0.9, 1.1)$. For perturbation, we set the probability $p$ of the Dropout layer as 0.1. For binary interpolation, we sample $\alpha$ from a Bernoulli distribution $B(p=0.25)$. For Gaussian scaling, we sample $\beta$ from a normal distribution $N(0,0.1)$. We set the training epoch as 30. All experiments are conducted on a GeForce RTX A6000 GPU. 

\section{Results}\label{sec:results}

\begin{table*}[ht]
\centering 
\small
\setlength{\tabcolsep}{2.3pt}
% \resizebox{\textwidth}{!}{%
\begin{tabular}{@{}lcccccccccccc@{}}
\toprule
\multirow{2}{*}{Model} & \multicolumn{2}{c}{Ruby} & \multicolumn{2}{c}{JavaScript} & \multicolumn{2}{c}{Go} & \multicolumn{2}{c}{Python} & \multicolumn{2}{c}{Java} & \multicolumn{2}{c}{PHP} \\ \cmidrule(l){2-13} 
 & Original & w/ RA & Original & w/ RA & Original & w/ RA & Original & w/ RA & Original & w/ RA & Original & w/ RA \\ \midrule
RoBERTa (code) & 0.641  & \textbf{0.665} & 0.583 & \textbf{0.612} & 0.867 & \textbf{0.892} & 0.610 & \textbf{0.663} & 0.634 & \textbf{0.674} & 0.584 & \textbf{0.617} \\
CodeBERT & 0.648 & \textbf{0.664} & 0.594 & \textbf{0.608} & 0.878 & \textbf{0.890} & 0.636  & \textbf{0.654} & 0.663 & \textbf{0.674} & 0.615 & \textbf{0.619} \\
GraphCodeBERT & 0.705 & \textbf{0.721} & 0.647 & \textbf{0.671} & 0.896 & \textbf{0.903} & 0.690 & \textbf{0.708} & 0.691 & \textbf{0.708} & 0.648  & \textbf{0.656} \\ \bottomrule
\end{tabular}%
% }
\caption{Performance of different approaches under MRR. ``w/ RA'' stands for ``with representation-level augmentation''.}
\label{overallresult}
\end{table*}

In this section, we first show the overall performance on code search when applying representation-level augmentation, and then individually analyze each augmentation method. Then, we demonstrate the relationship between loss and MRR and the effect of augmentation on vector distribution. Finally, we take an ablation study to analyze the impact of augmentation times. 

\subsection{Overall results}
The overall performance evaluation results are shown in Table \ref{overallresult}. In each iteration, we randomly choose one augmentation method. We can see that representation-level augmentation is a universal approach that can consistently improve the code search performance. 
Optimizing a tighter lower bound of mutual information brings about 2\% gain of MRR on average. The robust improvements have no relationships with certain models or certain programming languages.

\subsection{Effectiveness of individual augmentation approach}
To evaluate the effectiveness of the five augmentation approaches, we apply them alone and test them on the CSN-Python dataset, as shown in Table \ref{augalone}. Note that we follow the same experimental settings described in \Sec~\ref{sec:expsetup}. As the results show, every augmentation approach can improve baselines and the improvements brought by these augmentation approaches are stable. The combination of these approaches does not boost the performance compared with individually applying one of these augmentations. Since all these approaches can be derived from the general format, they have no distinct difference and hence share the similar effect on improving the mutual information between positive pairs.

\begin{table}[h]
\centering 
\footnotesize
\setlength{\tabcolsep}{1pt}
% \resizebox{\columnwidth}{!}{%
\begin{tabular}{@{}lccc@{}}
\toprule
Augmentations & RoBERTa & CodeBERT & GraphCodeBERT \\ \midrule
no augmentations & 0.629 & 0.636 & 0.690 \\ 
linear interpolation & 0.644 & 0.648 & 0.702 \\
linear extrapolation & 0.640 & 0.646 & 0.704 \\
stochastic perturbation & 0.658 & 0.648 & 0.698 \\
binary interpolation & 0.655 & \textbf{0.655} & 0.705 \\
Gaussian scaling  & 0.657 & 0.649 & 0.696 \\  
all augmentations  & \textbf{0.663}  & 0.654 & \textbf{0.708} \\ \bottomrule
\end{tabular}%
% }
\caption{Results of individual augmentation on CSN-Python dataset.}
\label{augalone}
\end{table}

\begin{table}[h]
\resizebox{\columnwidth}{!}{%
\begin{tabular}{@{}lc@{}}
\toprule
Augmentations & MRR \\ \midrule
UniXCoder & 0.721  \\
UniXCoder + all augmentations & 0.699  \\
UniXCoder - normalization & 0.708  \\
UniXCoder - normalization + all augmentations & \textbf{0.728}  \\ \bottomrule
\end{tabular}%
}
\caption{MRR of UniXCoder on CSN-Python under different settings. }
\label{limitation1}
\end{table}

\begin{figure}[h]
\centering
\includegraphics[width=\columnwidth]{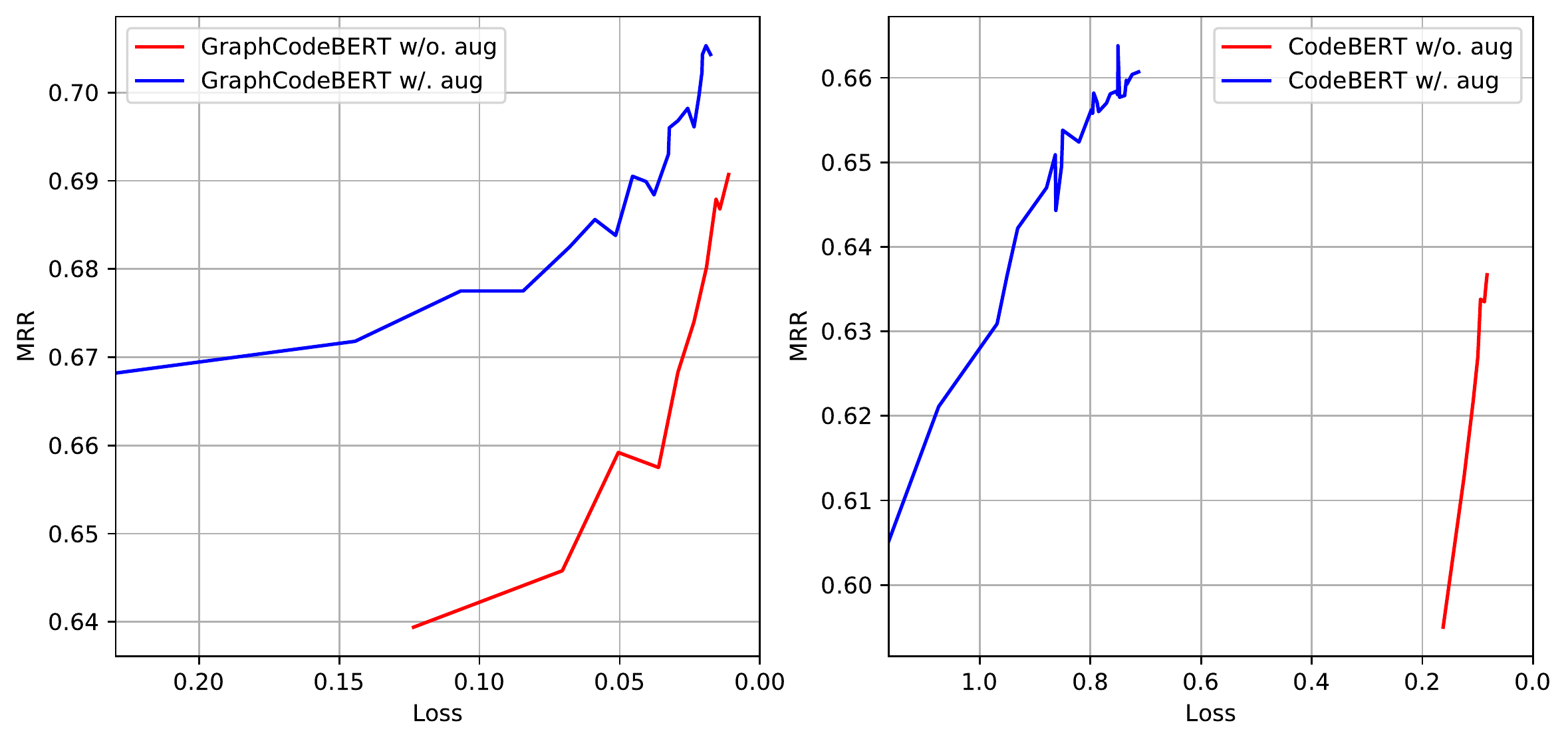}
\caption{The relationship between loss and MRR with or without representation-level augmentation. {\it aug} is short for representation-level augmentation.}
\label{loss-mrr}
\end{figure}

\begin{figure*}[h]
\centering
\includegraphics[width=\linewidth]{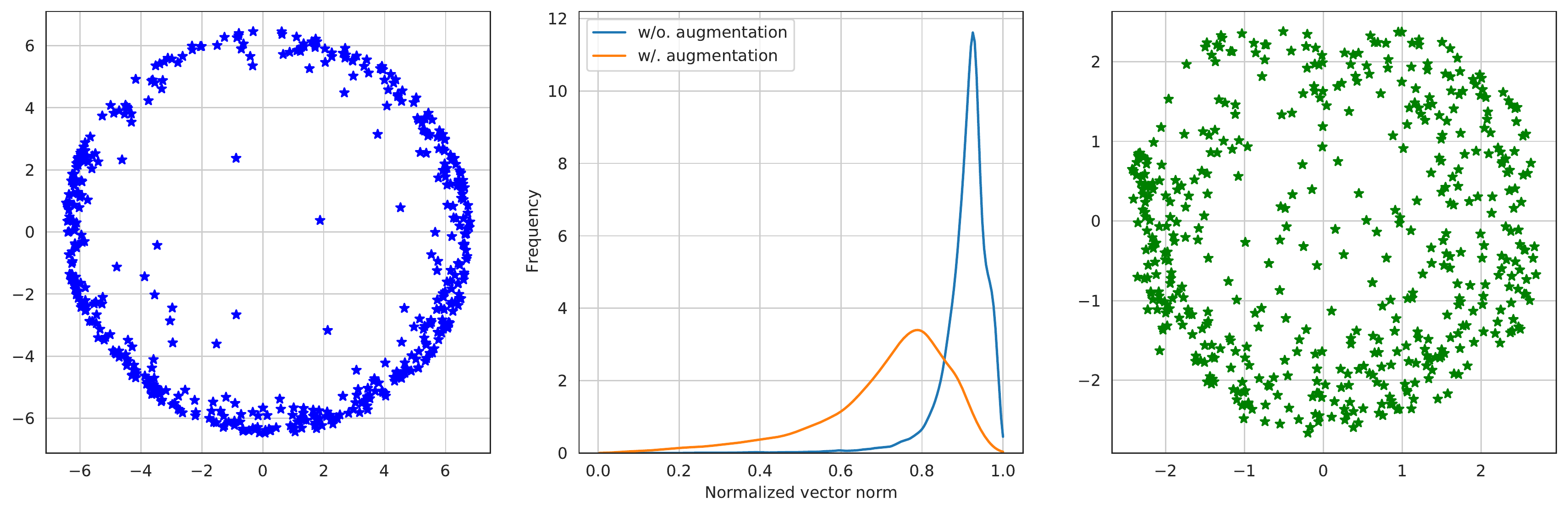}
\caption{Visualization of code vector distribution with and without representation-level augmentation. \textbf{Left}: without augmentation. \textbf{Right}: with augmentation. \textbf{Middle}: distribution of vector norms.}
\label{vectordist}
\end{figure*}

\subsection{Relationship between Loss and MRR}
In theoretical analysis, we get the conclusion that the relationship between the mutual information of positive pairs and loss is $I(q,c) \ge \log (B)-\mathcal{L_N}$ while for representation-level augmentation $I(q,c) \ge \frac{1}{\alpha^2}(\log (NB)-\mathcal{L_N})$. Thus, we argue that the effect of representation-level augmentation is improving the lower bounds of mutual information. However, this only comes true when loss can decrease to similar values under such two conditions. To prove that, after each epoch, we save the model parameters, record the loss of the training set, test models on the CSN-Python test set, and plot the relationship between loss and MRR with or without representation-level augmentation, as shown in Figure \ref{loss-mrr}. We can see that when the loss is the same, representation-level augmentation always leads to better performance (GraphCodeBERT). Even when loss cannot decrease to the same value without augmentation, it outperforms the vanilla contrastive learning (CodeBERT). We believe that other than improving the lower bounds, capturing the explicit relations between augmented data and the original one can also lead to higher mutual information between positive pairs.

\subsection{Impact on vector distribution}
\label{cos}
In code search, we measure the similarity by calculating the dot product between query representations and code representations. Here we analyze the influence of representation-level augmentation by visualizing the code vector distribution. We add a linear layer to embed the representations to a two-dimensional vector, as shown in Figure \ref{vectordist}. Besides, we plot the two-norms of vectors with Gaussian kernel density estimation. 

As \citet{wang2020understanding} concluded, the optimization of InfoNCE loss makes vectors evenly distributed, which corresponds to the left image of Figure \ref{vectordist}. Vectors roughly have the same two-norm. After applying augmentation, vectors still follow a circular pattern but their two-norms are changed. We argue that the InfoNCE loss degrades dot product to cosine distance since all vectors have similar norms while representation-level augmentation leverages the norms of vectors to distinguish some hard examples. We can see that in the middle image of Figure \ref{vectordist}, vector norms are much more uniformly distributed than that without augmentation.

However, this impact also reveals a limitation of representation-level augmentation. Some previous studies argue the necessity of using vector normalization, otherwise, the {\it Softmax} distribution will be made arbitrarily sharp. For example, the last step of UniXCoder is normalization. We evaluate UniXCoder on CSN-Python, as shown in Table \ref{limitation1}. When we apply augmentations to UniXCoder with normalization, the performance is worse. However, if we remove the normalization step, augmentations can boost performance just like for other models.

%\paragraph{Impact of augmentation times}
\subsection{Impact of the number of augmentation times}

\begin{figure}[t]
\centering
\includegraphics[width=0.85\columnwidth]{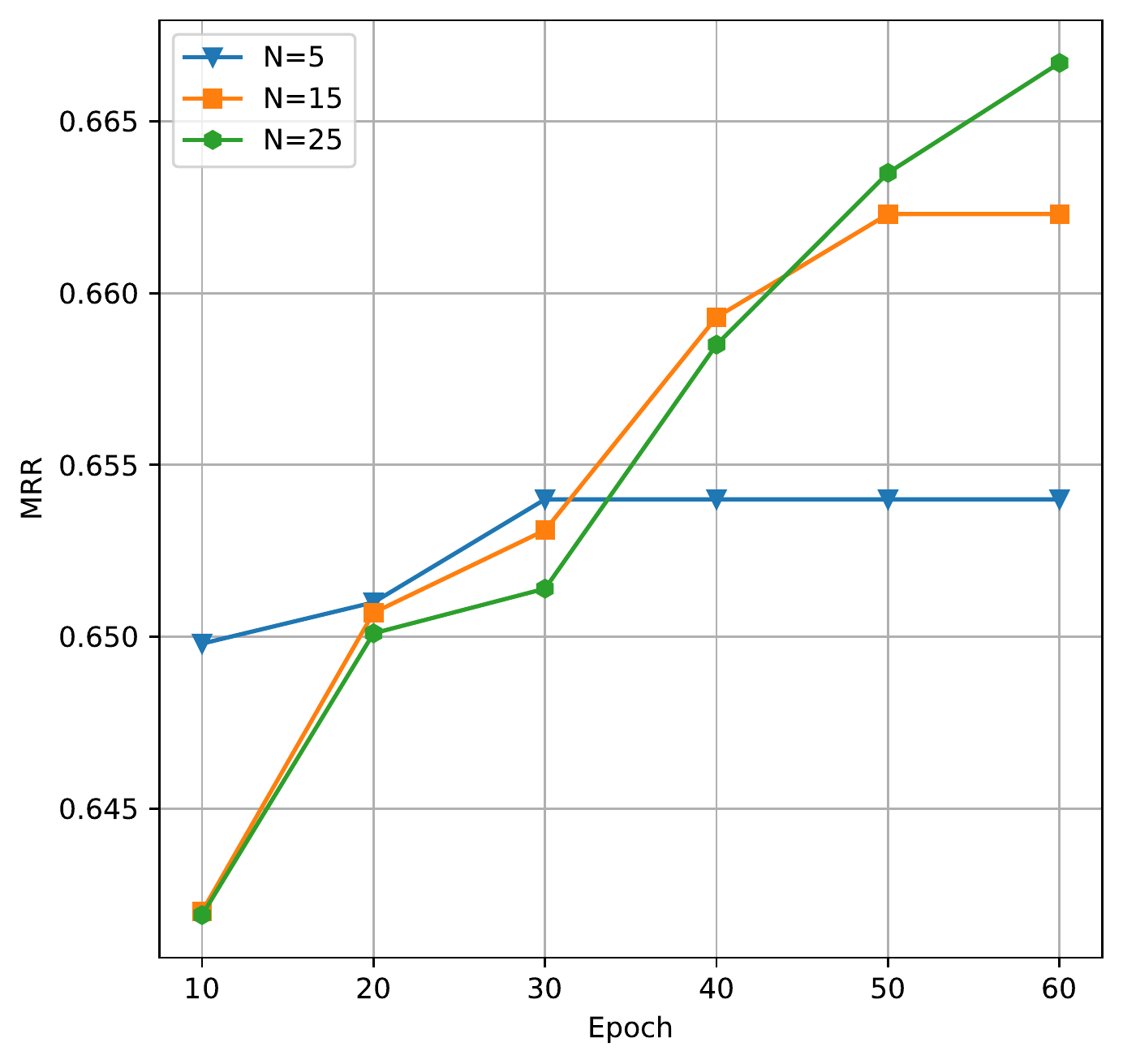}\caption{Performance of CodeBERT on each epoch interval with different augmentation times $N$.}
\label{mixtime}
\vspace{-6pt}
\end{figure}

To analyze the impact of augmentation times $N$, we conduct experiments on CodeBERT with $N=5,15,25$, respectively. We take 10 epochs as an interval, save the best model in each interval and test them on the test set, as shown in Figure \ref{mixtime}. We can see that augmenting more times leads to a relatively better performance that is even greater than the result reported in Table \ref{overallresult}. However, better results require more training time. According to $I(q,c) \ge \frac{1}{\alpha^2}(\log (NB)-\mathcal{L_N})$, bigger $N$ is better, but time cost of minimizing $\mathcal{L_N}$ also increases significantly. As we could see in the figure, a 5-times augmentation takes 30 epochs to converge while a 15-times augmentation takes 50 epochs. We train CodeBERT with $N=25$ for 60 epochs and MRR is still increasing. Therefore, we argue that for the application of representation-level augmentation, we should find a balance between performance and time cost.

\begin{table}[]
\resizebox{\columnwidth}{!}{%
\begin{tabular}{lcccc}
\toprule
\multirow{2}{*}{Model} & \multicolumn{2}{c}{FiQA-2018} & \multicolumn{2}{c}{NFCorpus} \\ \cline{2-5} 
 & Original & w/ RA & Original & w/ RA \\ \midrule
DistilBERT & 0.352 & \textbf{0.400} & 0.481 & \textbf{0.505} \\
RoBERTa & 0.343 & \textbf{0.356} & 0.367 & \textbf{0.389} \\ \bottomrule
\end{tabular}%
}
\caption{The performance of different approaches on MRR@1000. Here we follow the widely used metrics on passage retrieval tasks. MRR@1000 only considers the top 1000 returned passages.
}
\vspace{-6pt}
\label{passageretrieval}
\end{table}

\section{Discussion}\label{sec:discussion}
% \paragraph{Application on passage retrieval datasets}

According to the proof of Theorem \ref{theorem2}, the advantage of applying representation-level augmentation should be task-agnostic. To show its generalization ability, we evaluate our proposed approaches on two passage retrieval benchmark datasets, NFCorpus \cite{nfcorpus} and FiQA-2018\footnote{https://sites.google.com/view/fiqa/}. The difference between code search and passage retrieval is that the retrieved items are changed from code snippets written in programming language to passages written in English. We take DistilBERT \cite{sanh2019distilbert} and RoBERTa \cite{liu2019roberta} 
for experiments. We implement our approach based on an open-sourced framework BEIR \cite{beir}. The two models are fine-tuned on two datasets for 20 epochs, respectively. The settings of augmentation are the same as those in the code search task. For other hyper-parameters, we follow the settings that are provided by the framework. Results are shown in Table \ref{passageretrieval},  
which confirms that representation-level augmentation also improves the performance of passage retrieval models.

\section{Conclusion}\label{sec:conclusion}
In this work, we unify existing approaches to propose a general format of representation-level augmentation in code search. Based on the general format, we propose three other augmentation methods. We further theoretically analyze the effect of representation-level augmentation by proving that it helps optimize a tighter lower bound of mutual information between positive pairs. We evaluate our approach on several models and datasets and the results demonstrate the effectiveness of the proposed approach.

\section*{Limitations}\label{limitations}
As discussed in \Sec \ref{sec:results}, representation-level augmentation mainly has two limitations. First, it cannot boost the performance for models with vector normalization. We find that representation-level augmentation improves performance by leveraging the norms of vectors. With normalization, the norm of vectors is fixed and hence augmentation cannot bring performance gains. Second, as discussed in Section 6.5, although augmenting more times can lead to better performance, the time spent on training also increases. 
Balancing performance and time-cost will be our future work.

\section{Acknowledgement}
We thank the anonymous reviewers for their helpful comments and suggestions. This research is supported in part by the National Research Foundation, Prime Minister's Office, Singapore under its NRF Investigatorship Programme (NRFI Award No. NRF-NRFI05-2019-0002). Any opinions, findings and conclusions or recommendations expressed in this material are those of the authors and do not reflect the views of National Research Foundation, Singapore. This research is also supported in part by the China-Singapore International Joint Research Institute (CSIJRI), Guangzhou, China (Award No. 206-A021002) and the Joint NTU-WeBank Research Centre on Fintech (Award No. NWJ-2020-007), Nanyang Technological University, Singapore.

\normalem
\bibliographystyle{acl_natbib}
\bibliography{custom}

% \clearpage
\appendix

\section{Proof}
\subsection{Proof of Theorem 1}
\label{appendix:theory1}
\begin{small}
\begin{equation}
\begin{aligned}
    \mathcal{L_N} &= -\mathbb{E} \left [\log \frac{\exp (q_i\cdot c_i)}{\exp (q_i\cdot c_i)+\sum_{j\neq i}^B\exp (q_i \cdot c_j)} \right] \\
    &=\mathbb{E} \left [\log \left (1+\frac{\sum_{j\neq i}^B\exp (q_i \cdot c_j)}{\exp (q_i\cdot c_i)} \right ) \right] \\
    & \ge \mathbb{E} \left [\log \frac{\sum_{j\neq i}^B\exp (q_i \cdot c_j)}{\exp (q_i\cdot c_i)} \right] \\
    &=\mathbb{E} \left [ \log \left (\sum_{j\neq i}^B \exp (q_i \cdot c_j) \right) - \log \exp (q_i\cdot c_i) \right ] \\
    & \approx \mathbb{E} \Big[ \log \left [B\cdot \mathbb{E}[\exp (q_i \cdot c_j)]\right ] \Big] \\
    & \quad - \mathbb{E} \Big[ \log \exp (q_i \cdot c_i) \Big] \\
\end{aligned}
\end{equation}
\end{small}

According to the original paper of InfoNCE loss \cite{van2018representation}, the optimal value for $\exp (q \cdot c)$ is given by $\frac{p(c|q)}{p(c)}$, by substituting it, we can get:

\begin{equation}
\label{substitute}
\begin{aligned}
    \mathcal{L_N} & \ge \mathbb{E} \Big[ \log \left [B\cdot \mathbb{E}[\frac{p(c_j|q_i)}{p(c_j)})]\right ] \Big] \\
    & \quad - \mathbb{E} \Big[ \log \frac{p(c_i|q_i)}{p(c_i)} \Big] \\
    % &= \log B - I(q_i,c_i)
\end{aligned}
\end{equation}

Since $q_i$ and $c_j$ are negative pairs and sampled independently, $\mathbb{E}[\frac{p(c_j|q_i)}{p(c_j)}] = \mathbb{E} [\frac{p(c_j,q_i)}{p(c_j)p(q_i)}] = \mathbb{E} [\frac{p(c_j)p(q_i)}{p(c_j)p(q_i)}]=1$. According to the definition of mutual information described in Eq.\ref{mi}, the mutual information $I(q_i,c_i)$ is the second term. Thus, we get:

\begin{equation}
    \mathcal{L_N} \ge \log B - I(q_i,c_i)
\end{equation}

Therefore, $I(q,c) \ge \log (B)-\mathcal{L_N}$.

\subsection{Proof of Theorem 2}
\label{appendix:theory2}

By applying augmentation, the expectation of loss can be divided into four parts that correspond to four types of pairs: original query and original code, augmented query and original code, original query and augmented code, and augmented query and augmented code. It can be expressed as:

\begin{small}
\begin{equation}
\begin{aligned}
    \mathcal{L_N} & = \mathbb{E} \Bigg[-\mathbb{E} \left [\log \frac{\exp (q_i\cdot c_i)}{\exp (q_i\cdot c_i)+\sum_{j\neq i}^{BN}\exp (q_i \cdot c_j)} \right] \\
    & \quad - \mathbb{E} \left [\log \frac{\exp (q_i\cdot c_i^*)}{\exp (q_i\cdot c_i^*)+\sum_{j\neq i}^{BN}\exp (q_i \cdot c_j)} \right] \\
    & \quad - \mathbb{E} \left [\log \frac{\exp (q_i^*\cdot c_i)}{\exp (q_i^*\cdot c_i)+\sum_{j\neq i}^{BN}\exp (q_i^* \cdot c_j)} \right] \\ 
    & \quad - \mathbb{E} \left [\log \frac{\exp (q_i^*\cdot c_i^*)}{\exp (q_i^*\cdot c_i^*)+\sum_{j\neq i}^{BN}\exp (q_i^* \cdot c_j)} \right] 
    \Bigg]
\end{aligned}
\end{equation}
\label{loss}
\end{small}
where $q_i \in Q$, $c_i \in C$, $q_i^* \in Q^+$, $c_i^* \in C^+$, and $c_j \in C \cup C^+$. Next, we will analyze the four terms individually. For the first term, original query and original code, it is the same as proved in Appendix \ref{appendix:theory1}. For the second term, original query and augmented code, the derivation can be formulated as:

\begin{equation}
\begin{aligned}
    & \quad - \mathbb{E} \left [\log \frac{\exp (q_i\cdot c_i^*)}{\exp (q_i\cdot c_i^*)+\sum_{j\neq i}^{BN}\exp (q_i \cdot c_j)} \right] \\
    & =\mathbb{E} \left [\log \left (1+\frac{\sum_{j\neq i}^{BN}\exp (q_i \cdot c_j)}{\exp (q_i\cdot c_i^*)} \right ) \right] \\
    & \ge \mathbb{E} \left [\log \left (\frac{\sum_{j\neq i}^{BN}\exp (q_i \cdot c_j)}{\exp (q_i\cdot (\alpha \cdot c_i + \beta \cdot c_j))} \right ) \right] \\
    & = \mathbb{E} \left [\log \left (\frac{\sum_{j\neq i}^{BN}\exp (q_i \cdot c_j)}{\exp (q_i\cdot c_i\cdot \alpha) \cdot \exp(q_i\cdot c_j\cdot \beta))} \right ) \right] \\
    & = \mathbb{E} \Big[ \log \sum_{j\neq i}^{BN} \exp (q_i\cdot c_j)-\alpha \cdot \log \exp (q_i\cdot c_i) \\
    & \quad -\beta \cdot \log \exp (q_i \cdot c_j) \Big] \\
\end{aligned}
\end{equation}

Similar to Eq.\ref{substitute}, we substitute exponential function with probability, and we could get that the second term is greater than $\log NB -\alpha I(q_i, c_i)-\beta I(q_i,c_j)$. The only difference between the second term and the third term is that in the derivation of second term we decompose $c_i^*$ into $\alpha \cdot c_i + \beta \cdot c_j$ while for the third one we decompose $q_i^*$ into $\alpha \cdot q_i + \beta \cdot q_j$. Therefore, the third term is greater than $\log NB -\alpha I(q_i, c_i)-\beta I(q_j,c_i)$.

For the fourth term, it can be described as:

\begin{small}
\begin{equation}
\begin{aligned}
    & \quad - \mathbb{E} \Bigg[ \log \frac{\exp (q_i^*\cdot c_i^*)}{\exp (q_i^*\cdot c_i^*)+\sum_{j\neq i}^{BN}\exp (q_i^* \cdot c_j)} \Bigg] \\
    & \ge \mathbb{E} \Bigg[ \log \left (\frac{\sum_{j\neq i}^{BN}\exp (q_i^* \cdot c_j)}{\exp ((\alpha \cdot q_i + \beta \cdot q_j))\cdot (\alpha \cdot c_i + \beta \cdot c_j^\prime))} \right ) \Bigg] \\
    &= \mathbb{E} \Big[ \log \sum_{j\neq i}^{BN} \exp (q_i^*\cdot c_j)-\alpha^2 \cdot \log \exp (q_i\cdot c_i) \\
    & \quad -\alpha \beta \cdot \log \exp (q_i \cdot c_j^\prime) -\alpha \beta \cdot \log \exp (q_j \cdot c_i) \\
    & \quad -\beta^2 \cdot \log \exp (q_j \cdot c_j^\prime) \Big] \\
    & \ge \log NB -\alpha^2 I(q_i, c_i)-\alpha\beta I(q_i,c_j^\prime) \\
    & \quad -\alpha\beta I(q_j,c_i) - \beta^2 I(q_j,c_j^\prime) \\
\end{aligned}
\end{equation}
\end{small}

Then we remove the expectation of Eq.\ref{loss} by multiplying the corresponding probabilities of four terms, we get:

\begin{small}
\begin{equation}
\begin{aligned}
    \mathcal{L_N} & \ge \frac{B^2}{(N+1)^2B^2}\Big(\log NB - I(q_i,c_i)\Big) \\
    & \quad + \frac{NB^2}{(N+1)^2B^2}\Big(\log NB -\alpha I(q_i, c_i)-\beta I(q_i,c_j)\Big) \\
    & \quad + \frac{NB^2}{(N+1)^2B^2}\Big(\log NB -\alpha I(q_i, c_i)-\beta I(q_j,c_i)\Big) \\
    & \quad + \frac{N^2B^2}{(N+1)^2B^2}\Big(\log NB -\alpha^2 I(q_i, c_i)-\alpha\beta I(q_i,c_j^\prime) \\
    & \quad -\alpha\beta I(q_j,c_i) - \beta^2 I(q_j,c_j^\prime)\Big) \\
    & = \log NB - \Big( \frac{(\alpha N+1)^2}{(N+1)^2}I(q_i,c_i) \\ 
    &\quad + \frac{N\beta+N^2\alpha\beta}{(N+1)^2} I(q_i,c_j^\prime) + \frac{N\beta+N^2\alpha\beta}{(N+1)^2} I(q_j,c_i) \\
    & \quad + \frac{N^2\beta^2}{(N+1)^2} I(q_j, c_j^\prime) \Big) \\
    & \ge \log NB - \Big(\alpha^2 I(q_i,c_i) +\alpha\beta\cdot I(q_i,c_j^\prime) \\
    &\quad +\alpha\beta\cdot I(q_j,c_i) +\beta^2\cdot I(q_j, c_j^\prime) \Big)
\end{aligned}
\end{equation}
\end{small}

Therefore, we get:
\begin{equation}
\begin{aligned}
    I(q_i,c_i) \ge &  \frac{1}{\alpha^2} \Big( \log (NB)-\mathcal{L_N} \\
    & -\alpha\beta\cdot I(q_i,c_j^\prime)-\alpha\beta\cdot I(q_j,c_i) \\
    & -\beta^2\cdot I(q_j, c_j^\prime) \Big)
\end{aligned}
\end{equation}

% \section{Experimental Setups}
% \label{expdetail}

% \paragraph{Datasets}
% The statistics of the dataset are shown in Table \ref{datastat}. We follow the instruction \cite{guo2020graphcodebert} to filter out low-quality examples. The following data is filtered out:
% \begin{itemize}
%     \item Codes that can not be parsed into Abstract Syntax Trees.
%     \item Queries that are shorter than 3 or longer than 256.
%     \item Queries that contain irrelevant contents like a url.
%     \item Queries written in other languages instead of English or empty queries.
% \end{itemize}

% \paragraph{Evaluation Metrics}
% We evaluate our approach with a widely adopted metric MRR (Mean Reciprocal Rank). MRR is the average of reciprocal ranks of a true code fragment for a given query $Q$. It is calculated as:
% \begin{equation}
%     MRR=\frac{1}{|Q|}\sum_{i=1}^{|Q|}\frac{1}{Rank_i}
% \end{equation}
% where $Rank_i$ is the rank of the correct code fragement that is related to the i-th query.

% \paragraph{}

% \begin{table}[]
% \resizebox{\columnwidth}{!}{%
% \begin{tabular}{lcccc}
% \hline
% Language & Training & Validation & Test & Codebase \\ \hline
% Ruby & 24,927 & 1,400 & 1,261 & 4,360 \\
% JavaScript & 58,025 & 3,885 & 3,291 & 13,981 \\
% Go & 167,288 & 7,325 & 8,122 & 28,120 \\
% Python & 251,820 & 13,914 & 14,918 & 43,827 \\
% Java & 164,923 & 5,183 & 10,955 & 40,347 \\
% PHP & 241,241 & 12,982 & 14,014 & 52,660 \\ \hline
% \end{tabular}%
% }
% \caption{CodeSearchNet Dataset statistics}
% \label{datastat}
% \end{table}

\section{Representation-level augmentation on other loss functions}
\label{otherloss}

Here, we investigate whether representation-level augmentation also benefits other loss functions. We apply representation-level augmentation on triplet loss \cite{triplet} and logistic loss \cite{logistic}, because these two loss functions are also used prior to InfoNCE loss. Triplet loss learns to maximize the distances between negative pairs while minimizing the distances between positive pairs, which can be written as:

\begin{equation}
\begin{aligned}
    L_{triplet} = & -\frac{1}{N} \sum_{i=1}^N max(0, \\ 
    & ||q_i-c_i||_2^2 - ||q_i-c_{j,j\neq i}||_2^2 + \epsilon)
\end{aligned}
\end{equation}

And logistic loss is also called NCE loss, which can be described as:

\begin{equation}
\begin{aligned}
    L_{logistic} = & -\frac{1}{N} \sum_{i=1}^N \Big[ \log \sigma(q_i \cdot c_i) \\ 
    & - \frac{1}{N-1} \sum_{j\neq i} \log \sigma(q_i \cdot c_j) \Big]
\end{aligned}
\end{equation}

where definitions of variables follow Equation \ref{infonce}, $\sigma$ represents sigmoid function, and we set $\epsilon=5$ in our experiments.

We finetune CodeBERT and GraphCodeBERT with the two loss functions for 20 epochs on Ruby and Javascript dataset. Results of triplet loss and logistic loss are shown in Table \ref{triplet} and Table \ref{logistic}, respectively.

\begin{table}[h]
\centering 
\small
\setlength{\tabcolsep}{2.3pt}
% \resizebox{\textwidth}{!}{%
\begin{tabular}{@{}lcccc@{}}
\toprule
\multirow{2}{*}{Model} & \multicolumn{2}{c}{Ruby} & \multicolumn{2}{c}{JavaScript} \\ \cmidrule(l){2-5} 
 & Original & w/ RA & Original & w/ RA  \\ \midrule
CodeBERT & 0.594 & \textbf{0.610} & 0.525 & \textbf{0.533}  \\
GraphCodeBERT & 0.683 & \textbf{0.698} & 0.620 & \textbf{0.637}  \\ \bottomrule
\end{tabular}%
% }
\caption{Performance of triplet loss under MRR. ``w/ RA'' stands for ``with representation-level augmentation''.}
\label{triplet}
\end{table}

\begin{table}[h]
\centering 
\small
\setlength{\tabcolsep}{2.3pt}
% \resizebox{\textwidth}{!}{%
\begin{tabular}{@{}lcccc@{}}
\toprule
\multirow{2}{*}{Model} & \multicolumn{2}{c}{Ruby} & \multicolumn{2}{c}{JavaScript} \\ \cmidrule(l){2-5} 
 & Original & w/ RA & Original & w/ RA  \\ \midrule
CodeBERT & 0.513 & \textbf{0.522} & 0.503 & \textbf{0.513}  \\
GraphCodeBERT & 0.696 & \textbf{0.705} & 0.612 & \textbf{0.632}  \\ \bottomrule
\end{tabular}%
% }
\caption{Performance of logistic loss under MRR. ``w/ RA'' stands for ``with representation-level augmentation''.}
\label{logistic}
\end{table}

As we can see, representation-level augmentation can also improve performance when other loss functions are used. We believe this is because the augmentation makes the model more robust.

\end{document}